\documentclass[12pt]{article}
\usepackage{amsmath,amsfonts,epsf}

\renewcommand{\Re}{\operatorname{Re}}
\renewcommand{\Im}{\operatorname{Im}}
\newcommand{\Tr}{\operatorname{Tr}}
\newcommand{\sign}{\operatorname{sign}}
\newcommand{\dd}{\text{d}}
\newcommand{\rr}{\boldsymbol r}
\newcommand{\pp}{p_v}

\begin{document}
\noindent
~\vspace{2.0cm}

\centerline{\LARGE Spectral statistics in chaotic systems with 
                   a point interaction}
\vspace{2.0cm}
\centerline{\large Martin Sieber\footnote{E-mail: 
sieber@mpipks-dresden.mpg.de}}
\vspace{0.5cm}

\centerline{
Max-Planck-Institut f\"ur Physik komplexer Systeme,
N\"othnitzer Str.\ 38, 01187 Dresden, Germany}

\vspace{5.0cm}
\centerline{\bf Abstract}
\vspace{0.5cm}
We consider quantum systems with a chaotic classical limit that are
perturbed by a point-like scatterer.  The spectral form factor
$K(\tau)$ for these systems is evaluated semiclassically in terms of
periodic and diffractive orbits. It is shown for order $\tau^2$ and
$\tau^3$ that off-diagonal contributions to the form factor which
involve diffractive orbits cancel exactly the diagonal contributions
from diffractive orbits, implying that the perturbation by the
scatterer does not change the spectral statistic. We further show that
parametric spectral statistics for these systems are universal for
small changes of the strength of the scatterer.

\vspace{2.5cm}

\noindent PACS numbers: \\
\noindent 03.65.Sq ~ Semiclassical theories and applications. \\
\noindent 05.45.Mt ~ Semiclassical chaos (``quantum chaos'').

\newpage

\section{Introduction}

Semiclassical theories for spectral statistics have been developed
\cite{HO84,Ber85,BK96} to find an explanation for the observed
universality in energy spectra of quantum systems with a chaotic
classical limit, the agreement of correlations in energy spectra with
those between eigenvalues of random matrices \cite{Boh91}.  They are based
on semiclassical trace formulas that approximate the density of states
in terms of classical trajectories \cite{Gut90}. It has been shown by
these theories that in the asymptotic limit of long-range correlations
two-point correlation functions do coincide with those of random
matrix theory \cite{Ber85,BK96}. These results are based on mean
properties of periodic orbits \cite{HO84}. To go beyond the leading
asymptotic term requires information about correlations between
periodic orbits which are presently not available \cite{ADDKKSS93}.

One of the expectations, on basis of the random matrix hypothesis
\cite{BGS84}, is that a perturbation of a chaotic system should not
change the statistical distribution of the energy levels of the
quantum system, if it does not change the chaotic nature of the
classical motion.  In the present article we investigate, on the level
of the semiclassical approximation, whether the perturbation by a
point-like scatterer has this property.  One argument in favour of
this invariance is that the semiclassical approximation for the
density of states is not changed in leading order of $\hbar$ for this
perturbation. The influence of the scatterer is described
semiclassically by a certain class of trajectories, so-called
diffractive orbits that start from the scatterer and return to
it. They contribute to the density of states in higher order of
$\hbar$ than the leading order contribution from periodic orbits.

The present article is motivated by the observation in \cite{Sie99b}
that a scatterer could nevertheless have an influence on
spectral statistics. When spectral correlation functions are
calculated by using mean properties of diffractive orbits, the
so-called diagonal approximation, they show modifications which, in
general, do not vanish in the semiclassical limit ($\hbar \rightarrow
0$). In order that this does not lead to deviations from random matrix
statistics, these terms have to be cancelled by off-diagonal terms
which contain information about correlations between different
trajectories. As remarked above, the calculation of correlations
between trajectories is an unsolved problem in general systems. For
the diffractive orbits that describe the influence of a scatterer,
however, off-diagonal terms can be calculated explicitly. This is
done in the following sections. The results show that diagonal and
off-diagonal terms indeed cancel each other.  Furthermore, the results
can be used to investigate parametric spectral correlations, i.\,e.\
correlations between spectra of the system for different parameter
values, where the parameter is the strength of the scatterer.  It is
shown that the parametric spectral correlations are universal for
small changes of the parameter.

\section{The spectral form factor}

\begin{figure}
\begin{center}
\mbox{\epsfxsize8cm\epsfbox{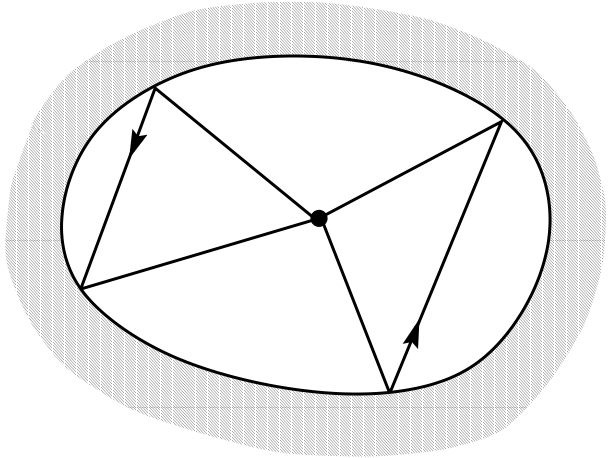}}
\end{center}
\vspace{0.0cm}
\caption{Example of a double-diffractive orbit.}
\label{examp}
\end{figure}

The perturbation by a point-like scatterer is represented, formally,
by a delta-potential $\hat{H} = \hat{H}_0 + \lambda \, \delta(\rr -
\rr_0)$, where $\lambda$ and $\rr_0$ are the strength and position of
the scatterer, respectively.  In more than one dimension such a
delta-potential is, however, not well defined. For example, it leads
to a divergent expression for the Green function. The problem can be
regularised by the method of self-adjoint extensions, leading to a
one-parameter family of Hamiltonians. A detailed monograph with
references on the history and on applications of delta-like
potentials is \cite{AGHH88}. We use in the following the property
that the semiclassical approximation for the density of states has
the same form as in the geometrical theory of diffraction
\cite{VWR94,PSc95,BW96,RVW96}, (see also \cite{DV98,Dah99}
for applications on spectral statistics).

We consider chaotic systems whose Hamiltonian is given in terms of
a scalar and a vector potential. Billiard systems can be included
in this description by letting the scalar potential be infinite
outside the billiard region. The statistical distribution of the
energy levels is investigated by semiclassically approximating
the spectral form factor. We restrict to two-dimensional systems
in order to keep the notation simple, but analogous calculations
can be performed in higher dimensions.

The spectral form factor is defined as Fourier transform of the
spectral two-point correlation function
\begin{equation} \label{koftau}
K(\tau) = \int_{-\infty}^{\infty} \! \frac{\dd \eta}{\bar{d}(E)} \;
\left\langle d_{\text{osc}}\left( E + \frac{\eta}{2} \right)
             d_{\text{osc}}\left( E - \frac{\eta}{2} \right)
\right\rangle_E
\; \exp\left( 2 \pi i \eta \tau \bar{d}(E) \right) \; .
\end{equation}
The function $d_{\text{osc}}(E)= d(E) - \bar{d}(E)$ is the
oscillatory part of the density of states, and $\bar{d}(E)$
is the smooth part which is given in two dimensions by 
$\bar{d}(E) \sim \Sigma(E) (2 \pi \hbar)^{-2}$, $E \rightarrow
\infty$, where $\Sigma(E)$ is the volume of the surface of
constant energy in phase space. The statistics is evaluated
by averaging over an energy interval that is small in comparison
to $E$ but contains a large number of energy levels. 

The semiclassical approximation for $K(\tau)$ is obtained by
inserting into (\ref{koftau}) the approximation for the
oscillatory part of the density of states
\begin{equation} \label{density}
d_{\text{osc}}(E) \approx \frac{1}{\pi \hbar} \Re \sum_\gamma A_\gamma
\exp\left( \frac{i}{\hbar} S_\gamma(E) \right) \; .
\end{equation}
In systems with a delta-like potential the sum in (\ref{density})
runs over all periodic orbits \cite{Gut90}, and further over all
diffractive orbits that start from the scatterer and return to
it an arbitrary number of times $n$ \cite{VWR94,PSc95,BW96,RVW96}.
An example for a double-diffractive orbit ($n=2$) in a billiard
system is shown in figure \ref{examp}. For $n$-fold diffractive
orbits the amplitude $A_\gamma$ has an $\hbar$-dependence of
$\hbar^{n/2}$, and $S_\gamma$ denotes the action of an orbit.

With (\ref{density}) one obtains the following approximation
for the spectral form factor
\begin{equation} \label{ksc}
K(\tau) = \frac{1}{2 \pi \hbar \bar{d}(E)} \left\langle 
\sum_{\gamma,\gamma'} A_{\gamma} A_{\gamma'}^*  \, 
\exp\left\{ \frac{i}{\hbar} \left( S_{\gamma}(E) - 
S_{\gamma'}(E) \right) \right\} 
\delta \left( T - \frac{T_{\gamma}+T_{\gamma'} }{2} \right) 
\right\rangle_E \; ,
\end{equation}
where $T = 2 \pi \hbar \bar{d}(E) \tau$, and $T_\gamma$ is the
period of an orbit. For small values of $\tau$
one can evaluate the double sum in (\ref{ksc}) in the diagonal
approximation \cite{Ber85}. One obtains in this way from the
periodic orbits the correct random matrix result
$K(\tau) \sim \frac{2}{\beta} \tau$,
$\tau \rightarrow 0$, where $\beta=1$ or $2$ for systems
with or without time-reversal symmetry, respectively. 

The diagonal contributions from diffractive orbits to the form
factor have been calculated in \cite{Sie99b}. The result for
$n$-fold diffractive orbits is
\begin{equation} \label{kdiag}
K_d^{(n)}(\tau)  = \frac{|{\cal D}|^{2n}}{(2 \beta)^n} \,
\frac{\tau^{n+1}}{n} \; ,
\end{equation}
where ${\cal D}$ is the diffraction coefficient for the
diffraction on the singularity of the potential \cite{AGHH88,ES96}.
It can be parameterised in the following form
\begin{equation} \label{d}
{\cal D} = \frac{2 \pi}{i \frac{\pi}{2} - \gamma - 
\log\left(\frac{k a}{2}\right)} \; .
\end{equation}
Here $k=\sqrt{2 m (E - V(\rr_0))}/\hbar$, $\rr_0$ is the
position of the scatterer, $a$ is a parameter describing
the strength of the potential, and $\gamma$ is Euler's constant.
In order that the terms (\ref{kdiag}) do not lead to a deviation
from random matrix statistics they have to be cancelled by
off-diagonal terms involving diffractive orbits. By calculating
off-diagonal terms for order $\tau^2$ and $\tau^3$ we show
in the following that such a cancellation does indeed occur.

We note that the diffraction coefficient satisfies the identities
\begin{equation} \label{did}
|{\cal D}|^2 = - 4 \Im {\cal D} \; , \qquad
|{\cal D}|^4 = 8( |{\cal D}|^2 - \Re {\cal D}^2) \; ,
\end{equation}
that will be used in the following. The first of these relations
expresses the conservation of probability, and the second is a
consequence of the first one.

\section{First-order correction}
\label{fo}

\begin{figure}
\begin{center}
\mbox{\epsfxsize8cm\epsfbox{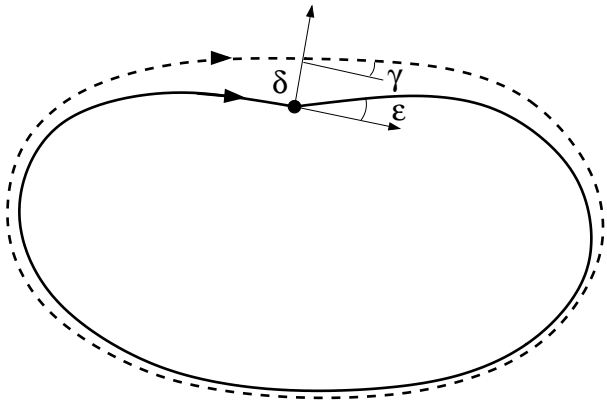}}
\end{center}
\vspace{0.0cm}
\caption{A diffractive orbit (full) that is almost periodic and a
         nearby periodic orbit (dashed). The local coordinate system
         is oriented along the final direction of the
         diffractive orbit.}
\label{figd1}
\end{figure}

The first-order correction to the diagonal approximation
for the form factor arises from off-diagonal terms
in (\ref{ksc}) between periodic orbits and single-diffractive
orbits. In leading order, these orbits are only correlated
if the diffractive orbit follows the periodic orbit very
closely. This happens, if the diffractive orbit is almost
periodic, i.\,e.\ if the final momentum is almost identical
to the initial momentum. An example is shown in figure \ref{figd1}.

The periodic orbit can be described by linearising
the motion around the diffractive orbit. The condition
that a trajectory in the vicinity of the diffractive 
orbit is periodic leads to the following equation
\begin{equation} \label{lind1}
\begin{pmatrix} \delta \\ \pp \gamma \end{pmatrix} 
= M \, 
\begin{pmatrix} \delta \\ \pp(\gamma - \varepsilon) \end{pmatrix} \; .
\end{equation}
Here $M$ is the stability matrix of the diffractive orbit,
$\varepsilon$ is the angle between the initial and final
direction of the diffractive orbit, $\gamma$ is the angle
between the direction along the periodic orbit and the final
direction of the diffractive orbit, and $\delta$ is the
spacial distance between periodic and diffractive orbit
(see figure \ref{figd1}). The quantity $\pp$ is defined
as mass times velocity, $\pp=mv$. The index $v$ is used
in order to distinguish it from the canonical momentum $p$
in systems with magnetic field. The stability matrix for
the motion in a magnetic field is discussed in the appendix.

In the linear approximation the difference in actions is
obtained by expanding the action up to second order
\begin{equation} \label{dsd1}
\Delta S(E) = S^{\text{po}}(E) - S^{\text{1do}}(E)
\approx - \pp \delta \varepsilon + \frac{1}{2}
(\Delta p_f - \Delta p_i) \delta
= - \frac{1}{2} \delta \varepsilon \pp \; ,
\end{equation}
where $\Delta p_f$ and $\Delta p_i$ are the differences between
the initial and final momenta of the periodic orbit and the
diffractive orbit, respectively. The solution of the
linear equation (\ref{lind1}) yields the following
relation between $\delta$ and $\varepsilon$
\begin{equation} \label{deld1}
\delta = \frac{M_{12}}{\Tr M - 2} \varepsilon \pp \; ,
\end{equation}
so that $\Delta S(E)$ depends quadratically on $\varepsilon$.

With this approximation the off-diagonal terms are
calculated. The amplitude of the diffractive orbit
is given by
\begin{equation} \label{a1do}
A_\gamma^{\text{1do}} = \frac{T_\gamma {\cal D}}{4 \pi \pp} 
\sqrt{\frac{2 \pi \hbar}{| (M_\gamma)_{12}|}}
\exp \left\{ - i \frac{\pi}{2} \nu_\gamma
- i \frac{3 \pi}{4} \right\} \; ,
\end{equation}
where $T_\gamma$ is the time along the orbit, $M_\gamma$
is its stability matrix, and $\nu_\gamma$ is the number
of conjugate points along the orbit. For the periodic
orbit the corresponding amplitude is
\begin{equation} \label{apo}
A_{\gamma}^{\text{po}} = \frac{T_{\gamma}}{\sqrt{| \Tr M_{\gamma}-2|}} 
\exp \left\{ - i \frac{\pi}{2} \mu_{\gamma} \right\} \; .
\end{equation}
The stability matrix is the same for both orbits in leading
order, but the Maslov index of the periodic orbit can differ
from the number of conjugate points $\nu_\gamma$ by 1 \cite{Gut90}
\begin{equation} \label{mu}
\mu_{\gamma} = \nu_{\gamma} + \frac{1}{2} 
(1 - \kappa_{\gamma}) \; ,
\qquad \kappa_{\gamma} = \sign \left( 
\frac{(M_{\gamma})_{12}}{\Tr M_{\gamma} - 2} \right) \; .
\end{equation}

In the following we sum over all diffractive orbits that
are almost periodic. This is done by applying first the
sum rule for diffractive orbits for which the angle
difference between initial and final direction has a
fixed value $\varepsilon$, and then integrating over the
angle $\varepsilon$. The sum rule is given by \cite{Sie99b}
\begin{equation} \label{sum1}
{\sum_{\gamma}}^{(\varepsilon)} \frac{1}{| (M_\gamma)_{12}|} \,
\delta(T - T_\gamma) \approx \frac{2 \pi \pp^2}{\Sigma(E)} \; ,
\end{equation}
where $\Sigma(E)$ is the volume of the energy shell. (It is
implied in (\ref{sum1}) that the left-hand side is smoothed
over small intervals of $T$ and $\varepsilon$ in order to
obtain a non-singular expression.)

Finally, one has to determine the multiplicity factor
of the contribution. First, each off-diagonal term
in (\ref{ksc}) has a corresponding complex conjugate partner.
If the summation is carried out over only one of these terms
one has to take twice the real part of the sum.
Furthermore, the periodic orbit and the diffractive orbit
both have multiplicities $2 \beta^{-1}$, but a particular
constellation occurs $2 \beta^{-1}$ times in the sum over
$\varepsilon$ (for systems with time-reversal symmetry
for $\varepsilon$ and $-\varepsilon$), so the total
multiplicity is $g=2 \beta^{-1}$.

Inserting the amplitudes (\ref{a1do}) and (\ref{apo}),
and the action difference (\ref{dsd1}) with (\ref{deld1})
into (\ref{ksc}) we obtain
\begin{align} \label{k1off}
K^{(1)}_{\text{off}}(\tau) & = \frac{g}{2 \pi \hbar \bar{d}(E)} \,
2 \Re \int_{-\infty}^\infty \! \dd \varepsilon
{\sum_{\gamma}}^{(\varepsilon)} A_{\gamma}^{\text{1do}}
(A_{\gamma}^{\text{po}})^*  \, 
\exp \left(- \frac{i}{\hbar} \Delta S_{\gamma}(E) \right)
\delta (T - T_\gamma) \notag \\ & =
\frac{4}{2 \pi \hbar \bar{d}(E) \beta}
\Re \int_{-\infty}^\infty \! \dd \varepsilon
{\sum_{\gamma}}^{(\varepsilon)} \frac{T_\gamma^2 \, {\cal D} \,
\sqrt{2 \pi \hbar} \, \delta (T - T_\gamma) 
}{4 \pi \pp \sqrt{|(M_\gamma)_{12} (\Tr M_\gamma - 2)|}}
\, e^{\frac{i \varepsilon^2 \pp^2}{2 \hbar} \,
\frac{(M_\gamma)_{12}}{\Tr M_\gamma - 2} - i \frac{\pi}{4}
(2 + \kappa_\gamma)} \notag \\ & = 
\frac{4}{2 \pi \hbar \bar{d}(E) \beta}
\Re \int_{-\infty}^\infty \! \dd \varepsilon'
{\sum_{\gamma}}^{(\varepsilon')} \frac{T_\gamma^2 \, 
{\cal D} \, \sqrt{2 \pi \hbar} \, \delta (T - T_\gamma) 
}{4 \pi \pp |(M_\gamma)_{12}|}
\, e^{- \frac{{\varepsilon'}^2 \pp^2}{2 \hbar}
- i \frac{\pi}{2}} \notag \\ & = 
\frac{2 T^2}{2 \pi \hbar \bar{d}(E) \beta}
\Im \int_{-\infty}^\infty \! \dd \varepsilon'
\frac{{\cal D} \, \pp \, \sqrt{2 \pi \hbar}}{\Sigma(E)}
\, e^{- \frac{{\varepsilon'}^2 \pp^2}{2 \hbar}} \notag \\ & =
\frac{2 \tau^2}{\beta} \Im {\cal D} \; .
\end{align}
The integration over $\varepsilon$ can be carried out from
minus to plus infinity since the main contribution comes
from the vicinity of $\varepsilon = 0$. Furthermore,
the following steps have been carried out. First the
integration variable has been changed to make the exponent
independent of the stability matrix.
Then the sum rule (\ref{sum1}) has been applied, assuming
that the distribution of angles between initial and final
momenta of a diffractive orbit is independent of the
distribution of the elements of the stability matrix
\begin{equation}
{\sum_{\gamma}}^{(\varepsilon)} \frac{1}{| (M_\gamma)_{12}|} \,
\delta(T - T_\gamma) \approx
{\sum_{\gamma}}^{(\varepsilon')} \frac{1}{| (M_\gamma)_{12}|} \,
\delta(T - T_\gamma) \; , \qquad
\varepsilon' = \varepsilon \sqrt{
\frac{- i (M_\gamma)_{12}}{\Tr M_\gamma - 2}}\; ,
\end{equation}
and finally the integration has been carried out.

$K^{(1)}_{\text{off}}(\tau)$ is the leading order correction
to the diagonal approximation for the form factor and it
cancels exactly the diagonal contribution from single
diffractive orbits ((\ref{kdiag}) with $n=1$). This can
be seen by using (\ref{did})
\begin{equation} \label{kd1}
K^{(1)}_{\text{d}}(\tau) + K^{(1)}_{\text{off}}(\tau) =
\frac{|{\cal D}|^2}{2 \beta} \tau^2 + \frac{2}{\beta} \tau^2
\Im {\cal D} = 0 \; .
\end{equation}
It shows that the presence of a point-like scatterer does
not modify the spectral form factor up to order $\tau^2$
in systems with a chaotic classical limit.

In order to find the geometries of orbits 
which contribute to a given order in $\tau$
it is helpful to count the orders of $\hbar$.
The $m$-th order off-diagonal correction to the
form factor is a $\tau^{m+1}$-term with a coefficient
that has to be $\hbar$-independent. 
The prefactor of the double sum over orbits in
(\ref{ksc}) is of order $\hbar^{-1}$ and the
product of the amplitudes of a $n_1$-fold and
a $n_2$-fold diffractive orbit is of order
$\hbar^{-(n_1+n_2)/2}$, where periodic orbits
are denoted here as 0-fold diffractive orbits. 
The conversion of time $T^{m+1}$ into $\tau^{m+1}$
gives an order $\hbar^{m+1}$, which yields altogether
an order of $\hbar^{(2m-n_1-n_2)/2}$. Furthermore,
every integration over a small parameter $\varepsilon$
gives an additional order $\hbar^{-1/2}$, if the
action difference is quadratic in this parameter.
As a consequence, $2m-n_1-n_2$ small parameters
are necessary in order that the prefactor of
$\tau^{(m+1)}$ is $\hbar$-independent. For the
first order correction in this section ($m=n_1=1$
and $n_2=0$) this estimate gives one small parameter
$\varepsilon$.

\section{Second-order corrections}

For the second-order corrections we consider orbits
that return twice to the region in coordinate space
from which they started. These orbits are close to
double-diffractive orbits. Double-diffractive
orbits have the semiclassical amplitude
\begin{equation} \label{a2do}
A_\gamma^{\text{2do}} = \frac{\hbar T_\gamma {\cal D}^2}{
16 \pi \pp^2} \frac{1}{\sqrt{|(R_\gamma)_{12} \, (L_\gamma)_{12}|}}
\exp \left\{ - i \frac{\pi}{2} (\nu_{\gamma,L} + \nu_{\gamma,R})
- i \frac{3 \pi}{2} \right\} \; ,
\end{equation}
where $T_\gamma$ is the total time along the
trajectory, $L_\gamma$ and $R_\gamma$ are the
stability matrices for the two loops
('left' and 'right'), and $\nu_{\gamma,L}$
and $\nu_{\gamma,R}$ are
the number of conjugate points along the loops.
(\ref{a2do}) is the amplitude for one particular
sequence in which the loops are traversed. In 
systems without time-reversal symmetry the
degeneracy of the trajectory is thus two,
meaning that there is another trajectory 
with exactly the same semiclassical amplitude
and action. This trajectory traverses first
the second loop of $\gamma$, and then the
first loop of $\gamma$. In systems with
time-reversal symmetry the degeneracy is eight.

The sum rule for double-diffractive orbits is given 
by \cite{Sie99b}
\begin{equation} \label{sum2}
\sum_\gamma^{(\varepsilon_1,\dots,\varepsilon_n)} 
\frac{1}{|(R_\gamma)_{12} \, (L_\gamma)_{12}|} \delta(T - T_\gamma) 
\approx \frac{(2 \pi \pp)^4}{\Sigma(E)^2} \frac{T}{(2 \pi)^n} \; ,
\end{equation}
if there are $n$ restrictions to the four directions
of the velocities at the point from which the trajectories
start and to which they return. As will be seen in the
following, it follows from this sum rule that the
contributions are of order $\tau^3$ (there is a factor $T$
from every semiclassical amplitude, and a factor $T$
from the sum rule).

There are several possibilities in which a double-diffractive
orbit can have an action which is almost identical to the
action of a single-diffractive or a periodic orbit. A
necessary condition is that there is always at least one
small relative angle between the different initial and
final directions of the orbit at the scattering point.
In order to find the relevant cases one has to consider
all possibilities and take into account the $\hbar$-argument
that was given at the end of section \ref{fo}. The result
is that there are three relevant configurations for systems
without time-reversal symmetry and five configurations for
systems with time-reversal symmetry. They are discussed
in the next sections.

\subsection{Correlations between double-diffractive and
single-diffractive orbits} \label{d2d1}

\begin{figure}
\begin{center}
\mbox{\epsfxsize10cm\epsfbox{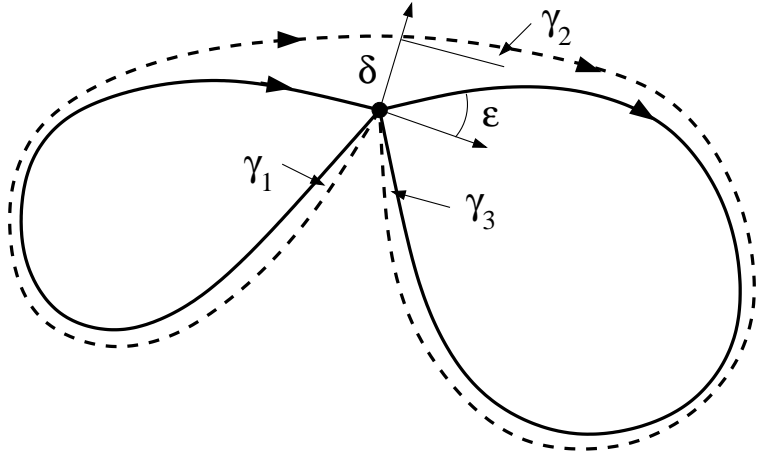}}
\end{center}
\vspace{0.0cm}
\caption{A double-diffractive orbit (full) for which one initial
         direction deviates by a small angle $\varepsilon$
         from one final direction, and a nearby single
         diffractive orbit (dashed).}
\label{figd2d1}
\end{figure}

Correlations between double-diffractive and single-diffractive
orbits exist if the double-diffractive orbit is almost
single-diffractive. This occurs if the final
velocity of one loop deviates by a small angle
$\varepsilon$ from the initial velocity of the other
loop. An example is shown in figure \ref{figd2d1}.
There is only one small parameter here which agrees
with the estimate $2m-n_1-n_2$ for $m=2$, $n_1=2$
and $n_2=1$.

The further calculations are done analogously to
the last section. The motion in the vicinity of the
double-diffractive orbit is linearised and one obtains
in this approximation a condition for the neighbouring
single-diffractive orbit
\begin{equation} \label{lind2d1}
\begin{pmatrix} \delta \\ \pp \gamma_2 \end{pmatrix} = L \, 
\begin{pmatrix} 0 \\ \pp \gamma_1 \end{pmatrix} \; , \qquad
\begin{pmatrix} 0 \\ \pp \gamma_3 \end{pmatrix} = R \, 
\begin{pmatrix} \delta \\ \pp (\gamma_2 - \varepsilon)
\end{pmatrix} \; .
\end{equation}
The angles $\gamma_1$, $\gamma_2$, $\gamma_3$ and the distance
$\delta$ are shown in figure \ref{figd2d1}. The difference
in action is obtained by expanding the action up to second
order 
\begin{equation} \label{dsd2d1}
\Delta S(E) = S^{\text{1do}}(E) - S^{\text{2do}}(E)
= - \frac{1}{2} \delta \varepsilon \pp
= - \frac{\varepsilon^2 \pp^2}{2} \frac{L_{12} \, R_{12}}{M_{12}} \; ,
\end{equation}
where $M=RL$ is the stability matrix of the single-diffractive
orbit. The last step in (\ref{dsd2d1}) follows from the solution
of (\ref{lind2d1}) for $\delta$.

The number of conjugate points $\nu_\gamma$ along the
single-diffractive orbit can differ from the sum of the
number of conjugate points along the two loops, $\nu_{\gamma,L}$
and $\nu_{\gamma,R}$, by 1. The general condition for this is
\begin{equation} \label{nu}
\nu_\gamma = \nu_{\gamma,L} + \nu_{\gamma,R}
+ \frac{1}{2}(1 - \sigma_\gamma) 
\; , \qquad \sigma_\gamma = \sign \left( \frac{(L_\gamma)_{12} 
\, (R_\gamma)_{12}}{(M_\gamma)_{12}} \right) \; .
\end{equation}

The single- and double-diffractive orbits have multiplicity
$2 \beta^{-1}$ and $8 \beta^{-2}$, respectively, but each
configuration occurs for $2 \beta^{-1}$ different values
of $\varepsilon$. Therefore the total multiplicity is
$g = 8 \beta^{-2}$. The contribution to the form factor
from all pairs of orbits is thus given by
\begin{align}
K^{(2a)}_{\text{off}}(\tau) & = \frac{g}{2 \pi \hbar \bar{d}(E)} \,
2 \Re \int_{-\infty}^\infty \! \dd \varepsilon
{\sum_{\gamma}}^{(\varepsilon)} A_{\gamma}^{\text{2do}} 
(A_{\gamma}^{\text{1do}})^*  \, 
\exp \left( - \frac{i}{\hbar} \Delta S_{\gamma}(E) \right)
\delta (T - T_\gamma) \notag \\ & =
\frac{16}{2 \pi \hbar \bar{d}(E) \beta^2}
\Re \int_{-\infty}^\infty \! \dd \varepsilon
{\sum_{\gamma}}^{(\varepsilon)} \frac{\hbar \, T_\gamma^2 \, 
{\cal D}^2 \, {\cal D}^* \, \sqrt{2 \pi \hbar} \, \delta (T - T_\gamma) 
}{64 \pi^2 \pp^3
\sqrt{| (L_\gamma)_{12} \, (R_\gamma)_{12} \, (M_\gamma)_{12} |}} 
\, e^{\frac{i \varepsilon^2 \pp^2}{2 \hbar} \,
\frac{(L_\gamma)_{12} 
\, (R_\gamma)_{12}}{(M_\gamma)_{12}} - i \frac{\pi}{4}
(2 + \sigma_\gamma)} \notag \\ & = 
\frac{1}{2 \pi \hbar \bar{d}(E) \beta^2} \Re
{\sum_{\gamma}}^{(\varepsilon')} \frac{\hbar^2 T^2 \, 
{\cal D}^2 \, {\cal D}^* \, \delta (T - T_\gamma) 
}{2 \pi \pp^4 |(L_\gamma)_{12} (R_\gamma)_{12}|}
\, e^{- i \frac{\pi}{2}} \notag \\ & = 
\frac{\tau^3}{\beta^2} \Im ({\cal D}^2 \, {\cal D}^*) \; .
\end{align}

Here we slightly abbreviated the procedure of (\ref{k1off})
and performed the integration directly.

\subsection{Correlations between double-diffractive orbits
and periodic orbits}

\begin{figure}
\begin{center}
\mbox{\epsfxsize12cm\epsfbox{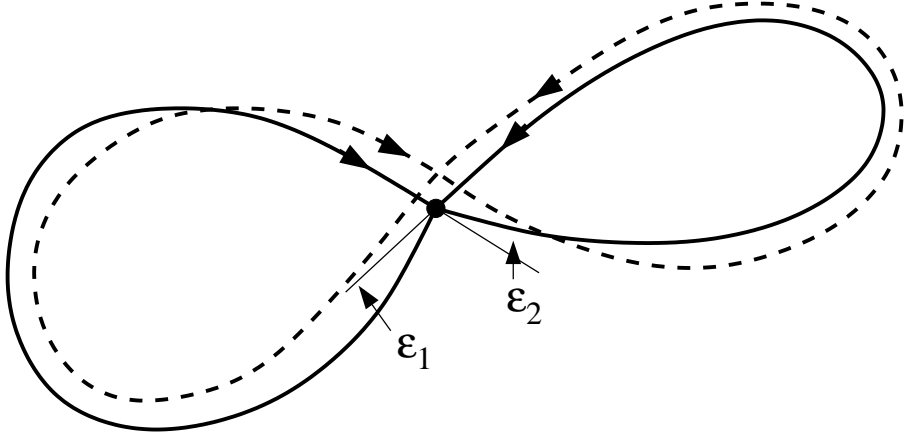}}
\end{center}
\vspace{0.0cm}
\caption{A double-diffractive orbit (full) for which the two initial
         directions deviate by small angles $\varepsilon_1$
         and $\varepsilon_2$ from the two final directions,
         and a nearby periodic orbit (dashed).}
\label{figd2po}
\end{figure}

In order that a double-diffractive orbit is close to a 
periodic orbit it has to be almost periodic. This means
that the final direction of each loop has to be almost
identical to the initial direction of the other loop,
so there are two small relative angles as is shown in
figure \ref{figd2po}.

The linearisation of the motion in the vicinity of the
double-diffractive orbit leads to the following condition
for the periodic orbit
\begin{equation} \label{lind2po}
\begin{pmatrix} \delta_2 \\ \pp \gamma_2 \end{pmatrix} = L \, 
\begin{pmatrix} \delta_1 \\ \pp (\gamma_1 - \varepsilon_1)
\end{pmatrix} \; , \qquad
\begin{pmatrix} \delta_1 \\ \pp \gamma_1 \end{pmatrix} = R \, 
\begin{pmatrix} \delta_2 \\ \pp (\gamma_2 - \varepsilon_2)
\end{pmatrix} \; .
\end{equation}
The angles $\gamma_i$ and distances $\delta_i$ are defined
analogously as before in terms of the local coordinate systems
that are oriented along the two final directions of the
diffractive orbit.

The difference in actions is given by
\begin{align} \label{dsd2po}
\Delta S(E) &= S^{\text{po}}(E) - S^{\text{2do}}(E) \notag \\ &
= - \frac{\pp}{2} (\delta_1 \varepsilon_1 + \delta_2 \varepsilon_2)
\notag \\ & = - \frac{\pp^2}{2} \,
\frac{(RL)_{12} \, \varepsilon_1^2 + (LR)_{12} \, \varepsilon_2^2
+ 2(L_{12} + R_{12}) \, \varepsilon_1 \varepsilon_2}{\Tr M - 2} \; ,
\end{align}
where $M=RL$ is the stability matrix for the periodic orbit. 
Equation (\ref{dsd2po}) can be written in the terms of a 
symmetric matrix $A$ such that 
$\Delta S(E) = - \frac{1}{2} \pp^2 \sum_{i,j} A_{ij} 
\varepsilon_i \varepsilon_j$. The number of negative 
eigenvalues of $A$ is given by the number of sign changes
in the sequence of sub-determinants ($1, A_{11}, \det A$).
With $A_{11} = M_{12}/(\Tr M - 2)$ and $\det A = 
L_{12} \, R_{12}/ (\Tr M - 2)$ one finds that the two signs
of the eigenvalues are given by
\begin{equation} \label{kaprho}
\kappa = \sign \left( \frac{M_{12}}{\Tr M - 2} \right) \, , \quad
\sigma = \sign \left( \frac{L_{12} \, R_{12}}{M_{12}} \right) \; .
\end{equation}

The Maslov index of the periodic orbit can now differ by zero,
one or two from the sum of conjugate points along the left and
right loop. The criterion for this is the combination of (\ref{mu})
and (\ref{nu}) and has the form
\begin{equation}
\mu = \nu_L + \nu_R + \frac{1}{2}(2 - \kappa - \sigma) \, ,
\end{equation}
where $\kappa$ and $\sigma$ are given in (\ref{kaprho}).

The multiplicities of double-diffractive and periodic orbits
are $8 \beta^{-2}$ and $2 \beta^{-1}$, respectively, but a
particular configuration of them occurs $4 \beta^{-1}$ times
in the integral over the angles (for example, for systems without
time reversal symmetry the two angles can be interchanged),
so the total multiplicity is $g = 4 \beta^{-2}$.

After inserting the amplitudes (\ref{a2do}) and (\ref{apo}) and
the action difference (\ref{dsd2po}) into (\ref{ksc}) we obtain
\begin{align}
K^{(2b)}_{\text{off}}(\tau) & = \frac{g}{2 \pi \hbar \bar{d}(E)} \,
2 \Re \int_{-\infty}^\infty \!
\dd \varepsilon_1 \, \dd \varepsilon_2
{\sum_{\gamma}}^{(\varepsilon_1,\varepsilon_2)} A_{\gamma}^{\text{2do}} 
(A_{\gamma}^{\text{po}})^*  \, 
\exp \left( - \frac{i}{\hbar} \Delta S_{\gamma}(E) \right)
\delta (T - T_\gamma) \notag \\ & =
\frac{8}{2 \pi \hbar \bar{d}(E) \beta^2}
\Re \int_{-\infty}^\infty \! \dd \varepsilon_1 \, \dd \varepsilon_2
{\sum_{\gamma}}^{(\varepsilon_1,\varepsilon_2)} 
\frac{\hbar \, T_\gamma^2 \, 
{\cal D}^2 \, \delta (T - T_\gamma) 
\, e^{ \frac{i \pp^2}{2 \hbar} \sum A_{i,j} 
\varepsilon_i \varepsilon_j
- i \frac{\pi}{4} (4 + \sigma_\gamma + \kappa_\gamma)}
}{16 \pi \pp^2 \sqrt{| (L_\gamma)_{12} \, (R_\gamma)_{12} \, 
(\Tr M_\gamma - 2) |}} \notag \\ & = 
\frac{1}{2 \pi \hbar \bar{d}(E) \beta^2} \Re
{\sum_{\gamma}}^{(\varepsilon_1',\varepsilon_2')}
\frac{\hbar^2 T^2 \, {\cal D}^2 \, \delta (T - T_\gamma) 
}{\pp^4 |(L_\gamma)_{12} (R_\gamma)_{12}|}
\, e^{- i \pi} \notag \\ & = 
- \frac{\tau^3}{\beta^2} \Re ({\cal D}^2) \; .
\end{align}

\subsection{Correlations between pairs of single-diffractive orbits}

For exactly the same kind of double-diffractive orbits as in
the last section, there is one further type of correlation
that has to be considered. It occurs because there are two
possible ways in which the double-diffractive orbit can be
deformed into a nearby single-diffractive orbit, and
consequently, there are correlations between these
single-diffractive orbits.

\begin{figure}
\begin{center}
\mbox{\epsfxsize12cm\epsfbox{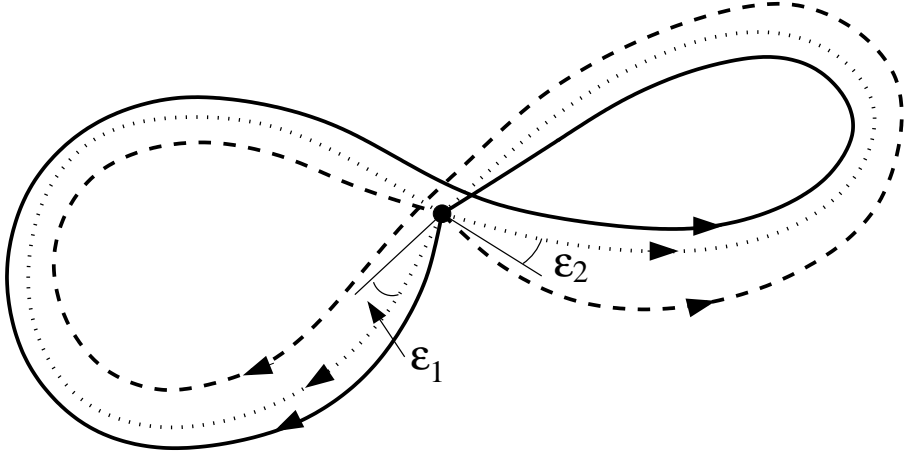}}
\end{center}
\vspace{0.0cm}
\caption{A double-diffractive orbit (dotted) for which the two initial
         directions deviate by small angles $\varepsilon_1$
         and $\varepsilon_2$ from the two final directions,
         and two nearby single-diffractive orbits (full and
         dashed).}
\label{figd1d1}
\end{figure}

The action difference between the two orbits can be
obtained from the action difference between each of these
orbits and the double-diffractive orbit (\ref{dsd2d1})
\begin{equation} \label{dsd1d1}
\Delta S(E) = S^{\text{1do,1}}(E) - S^{\text{1do,2}}(E)
= - \frac{1}{2} \delta_1 \varepsilon_1 \pp
  + \frac{1}{2} \delta_2 \varepsilon_2 \pp 
= - \frac{\pp^2}{2} \left( \frac{L_{12} \, R_{12}}{N_{12}} 
\varepsilon_1^2 - \frac{L_{12} \, R_{12}}{M_{12}} \varepsilon_2^2
\right) \; ,
\end{equation}
where $M=RL$ and $N=LR$ are the stability matrices of the two
single-diffractive orbits. The number of conjugate points along
the orbits are given by
\begin{alignat}{2} \label{nud1d1}
& \nu_1 = \nu_L + \nu_R + \frac{1}{2}(1 - \sigma_1) \; , \qquad &&
\nu_2 = \nu_L + \nu_R + \frac{1}{2}(1 - \sigma_2) \; , \notag \\ &
\sigma_1 = \sign \left( \frac{L_{12} \, R_{12}}{N_{12}} \right) 
\; , &&
\sigma_2 = \sign \left( \frac{L_{12} \, R_{12}}{M_{12}} \right) \; .
\end{alignat}

The multiplicities of the two orbits is both $2 \beta^{-1}$ and,
as before, the configuration occurs $4 \beta^{-1}$ times in the
double integral over the angles, so the total multiplicity is
$g = \beta^{-1}$.

Inserting the amplitude (\ref{a1do}) with stability matrix
$M$ and $N$, respectively, and the action difference
(\ref{dsd1d1}) into (\ref{ksc}) one obtains
\begin{align}
K^{(2c)}_{\text{off}}(\tau) & = \frac{g}{2 \pi \hbar \bar{d}(E)} \,
2 \Re \int_{-\infty}^\infty \! \dd \varepsilon_1 \dd \varepsilon_2
{\sum_{\gamma}}^{(\varepsilon_1,\varepsilon_2)}
A_{\gamma}^{\text{1do,2}}  (A_{\gamma}^{\text{1do,1}})^*  \, 
\exp \left( - \frac{i}{\hbar} \Delta S_{\gamma}(E) \right)
\delta (T - T_\gamma) \notag \\ & =
\frac{2}{2 \pi \hbar \bar{d}(E) \beta}
\Re \int_{-\infty}^\infty \! \dd \varepsilon_1 \, \dd \varepsilon_2
{\sum_{\gamma}}^{(\varepsilon_1,\varepsilon_2)}
\frac{2 \pi \hbar \, T_\gamma^2 \, 
|{\cal D}|^2 \, \delta (T - T_\gamma) 
\, e^{-\frac{i}{\hbar} \Delta S_\gamma(E) - 
i \frac{\pi}{4}(\sigma_{\gamma,1} - \sigma_{\gamma,2})} 
}{16 \pi^2  \pp^2 \sqrt{| (M_\gamma)_{12} \, (N_\gamma)_{12} |}} 
\notag \\ & = 
\frac{1}{2 \pi \hbar \bar{d}(E) \beta} \Re
{\sum_{\gamma}}^{(\varepsilon_1',\varepsilon_2')}
\frac{\hbar^2 T^2 \, |{\cal D}|^2 \, \delta (T - T_\gamma) 
}{2 \pp^4 |(L_\gamma)_{12} (R_\gamma)_{12}|} \notag \\ & = 
\frac{\tau^3}{2 \beta} |{\cal D}|^2 \; .
\end{align}

The contributions $K^{(2a)}_{\text{off}}(\tau)$ 
$K^{(2b)}_{\text{off}}(\tau)$ and $K^{(2c)}_{\text{off}}(\tau)$
are the only second-order off-diagonal corrections in
systems without time-reversal symmetry. As will be shown
in the following, these contributions cancel exactly
the diagonal term $K^{(2)}_{\text{d}}(\tau)$ for
$\beta=2$. For systems with time-reversal symmetry
there are further contributions. They arise from
the possibility that one trajectory can follow
one loop of another trajectory in the same direction,
but the other loop in the time-reversed direction.
We assume in the following that the relevant contributions
come from orbits which are close to double diffractive
orbits in coordinate space and we evaluate their
contributions in the next two subsections.

\subsection{Correlations between pairs of single-diffractive
orbits involving time-reversed loops}

\begin{figure}
\begin{center}
\mbox{\epsfxsize12cm\epsfbox{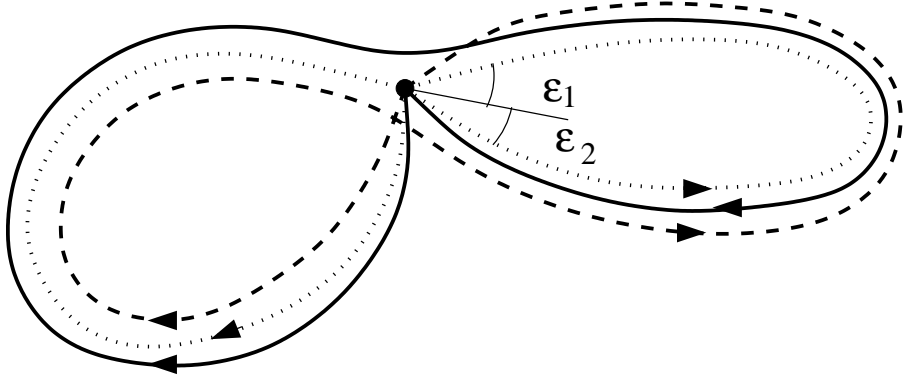}}
\end{center}
\vspace{0.0cm}
\caption{A double-diffractive orbit (dotted) with one loop for which
         the initial direction and the reversed final direction
         deviate by small angles $\varepsilon_1$ and $\varepsilon_2$
         from the final direction of the other loop. Furthermore,
         two nearby single-diffractive orbits (full and dashed),
         one of which traverses the second loop in the opposite
         direction (full).}
\label{figd1d1s}
\end{figure}

The first possibility involves two single-diffractive
orbits. These orbits follow closely a double-diffractive
orbit and traverse one loop in the same direction and the
other loop in the opposite direction. In order that this
can occur there must be one loop which has a very small
opening angle, and it must be almost aligned to the final
direction of the other loop as shown in figure \ref{figd1d1s}.

The action difference and the indices are given 
by the equations (\ref{dsd1d1}) and (\ref{nud1d1}) but now
with $M=RL$ and $N=R^i L$ where $R^i$ is the stability
matrix for the time-reversed second loop. In term of the
elements of $R$ the matrix $R^i$ is given by
\begin{equation}
R^i = \begin{pmatrix} R_{22} & R_{12} \\ R_{21} & R_{11}
      \end{pmatrix} \; .
\end{equation}

The multiplicity of each orbit is two, and the configuration
occurs two times in the integral over the angles, so the
total multiplicity is $g = 2$. Inserting the amplitude
(\ref{a1do}) with stability matrix $M$ and $N$, respectively,
and the action difference (\ref{dsd1d1}) into (\ref{ksc})
results in
\begin{align}
K^{(2d)}_{\text{off}}(\tau) & = \frac{g}{2 \pi \hbar \bar{d}(E)} \,
2 \Re \int_{-\infty}^\infty \! \dd \varepsilon_1 \dd \varepsilon_2
{\sum_{\gamma}}^{(\varepsilon_1,\varepsilon_2)}
A_{\gamma}^{\text{1do,2}}  (A_{\gamma}^{\text{1do,1}})^*  \, 
\exp \left( - \frac{i}{\hbar} \Delta S_{\gamma}(E) \right)
\delta (T - T_\gamma) \notag \\ & =
\frac{4}{2 \pi \hbar \bar{d}(E)}
\Re \int_{-\infty}^\infty \! \dd \varepsilon_1 \, \dd \varepsilon_2
{\sum_{\gamma}}^{(\varepsilon_1,\varepsilon_2)}
\frac{2 \pi \hbar \, T_\gamma^2 \, 
|{\cal D}|^2 \, \delta (T - T_\gamma) 
\, e^{-\frac{i}{\hbar} \Delta S_\gamma(E) 
- i \frac{\pi}{4} (\sigma_{\gamma,1} - \sigma_{\gamma,2})} 
}{16 \pi^2  \pp^2 \sqrt{| (M_\gamma)_{12} \, (N_\gamma)_{12} |}} 
\notag \\ & = 
\frac{1}{2 \pi \hbar \bar{d}(E)} \Re
{\sum_{\gamma}}^{(\varepsilon_1',\varepsilon_2')}
\frac{\hbar^2 T^2 \, |{\cal D}|^2 \, \delta (T - T_\gamma) 
}{\pp^4 |(L_\gamma)_{12} (R_\gamma)_{12}|} \notag \\ & = 
\tau^3 |{\cal D}|^2 \; .
\end{align}

\subsection{Correlations between single-diffractive orbits
and periodic orbits involving time-reversed loops}

\begin{figure}
\begin{center}
\mbox{\epsfxsize12cm\epsfbox{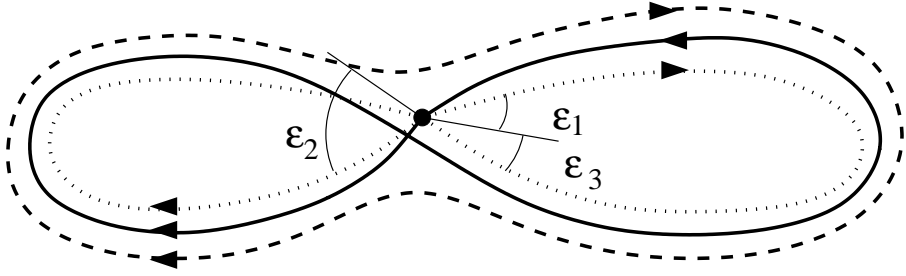}}
\end{center}
\vspace{0.0cm}
\caption{A double-diffractive orbit (dotted) for which
         all initial and final directions lie almost
         in one line. Furthermore, a nearby periodic orbit
         (dashed), and a nearby single-diffractive orbit
         (full) which traverses the second loop in the
         opposite direction.}
\label{figd1pos}
\end{figure}

The last relevant configuration occurs if all initial and final
velocities of the double-diffractive orbit lie almost in one
line as in figure \ref{figd1pos}. Then there exist neighbouring
single-diffractive and periodic orbits which follow one loop
in the same direction and the other loop in the opposite 
direction.

The action difference between the periodic orbit and
the diffractive orbit is obtained from (\ref{dsd2d1})
and (\ref{dsd2po})
\begin{align} \label{dsd1pos}
\Delta S(E) &= S^{\text{po}}(E) - S^{\text{1do}}(E) \notag \\ &
= - \frac{\pp}{2} (\delta_1 \varepsilon_1 + \delta_2 \varepsilon_2) 
+ \frac{\pp}{2} \delta_3 \varepsilon_3 
\notag \\ &
= - \frac{\pp^2}{2} \,
\frac{(RL)_{12} \, \varepsilon_1^2 + (LR)_{12} \, \varepsilon_2^2
+ 2(L_{12} + R_{12}) \, \varepsilon_1 \varepsilon_2}{\Tr M - 2} 
+ \frac{1}{2} \varepsilon_3^2 \pp^2 \frac{L_{12} \, R_{12}}{N_{12}} \, ,
\end{align}
where $M=RL$ and $N=R^i L$ are the stability matrices of the
periodic and diffractive orbit, respectively, and the indices
$\nu$ and $\mu$ of the diffractive and the periodic orbit are
\begin{gather}
\nu = \nu_L + \nu_R + \frac{1}{2}(1 - \sigma_2) \, , \qquad
\mu = \nu_L + \nu_R + \frac{1}{2}(2 - \kappa - \sigma_1) \, ,
\notag \\ 
\sigma_2 = \sign \left( \frac{L_{12} \, R_{12}}{N_{12}} \right)
\, , \quad
\kappa   = \sign \left( \frac{M_{12}}{\Tr M - 2} \right) \, , \quad
\sigma_1 = \sign \left( \frac{L_{12} \, R_{12}}{M_{12}} \right) \; .
\end{gather}

For each double-diffractive orbit like the one in figure
\ref{figd1pos} there are two periodic orbits of multiplicity
two and four single-diffractive orbits of multiplicity two
in the vicinity, but only half of the possible pairs involve
exactly one time reversed loop which makes a total of 16.
Furthermore, the double-diffractive orbit occurs 8 times
in the integral over the angles and the total multiplicity
is thus $g=2$.

Inserting the amplitudes (\ref{a1do}) and (\ref{apo}) and
the action difference (\ref{dsd1pos}) into (\ref{ksc}) yields
\begin{align}
K^{(2e)}_{\text{off}}(\tau) & = \frac{g}{2 \pi \hbar \bar{d}(E)} \,
2 \Re \int_{-\infty}^\infty \! \dd \varepsilon_1 \dd \varepsilon_2
\dd \varepsilon_3
{\sum_{\gamma}}^{(\varepsilon_1,\varepsilon_2,\varepsilon_3)}
A_{\gamma}^{\text{1do}} (A_{\gamma}^{\text{po}})^*  \, 
\exp \left(- \frac{i}{\hbar} \Delta S_{\gamma}(E) \right)
\delta (T - T_\gamma) \notag \\ & =
\frac{4}{2 \pi \hbar \bar{d}(E)}
\Re \int_{-\infty}^\infty \! \! \! \! \!
\dd \varepsilon_1 \dd \varepsilon_2 \dd \varepsilon_3
{\sum_{\gamma}}^{(\varepsilon_1,\varepsilon_2,\varepsilon_3)} 
\frac{T_\gamma^2 \, {\cal D} \,
\sqrt{2 \pi \hbar} \, \delta (T - T_\gamma) \,
e^{- \frac{i}{\hbar} \Delta S_{\gamma}(E)
-i \frac{\pi}{4} (2 - \sigma_{\gamma,2} + 
\kappa_{\gamma} + \sigma_{\gamma,1})} 
}{4 \pi \pp \sqrt{|(M_\gamma)_{12} (\Tr M_\gamma - 2)|}}
\notag \\ & = 
\frac{1}{2 \pi \hbar \bar{d}(E)}
\Re {\sum_{\gamma}}^{(\varepsilon_1',\varepsilon_2',\varepsilon_3')}
\frac{4 \pi \hbar^2 T^2 \, 
{\cal D} \, \delta (T - T_\gamma) 
}{\pp^4 |(L_\gamma)_{12} (R_\gamma)_{12}|} 
\, e^{-i \frac{\pi}{2}}   \notag \\ & =
2 \tau^3 \Im {\cal D} \; .
\end{align}

The sum of all contributions can be written in the form
\begin{align} \label{kd2}
K^{(2)}(\tau) & = K^{(2)}_{\text{d}}(\tau) + K^{(2a)}_{\text{off}}(\tau)
+ K^{(2b)}_{\text{off}}(\tau) + K^{(2c)}_{\text{off}}(\tau) +
K^{(2d)}_{\text{off}}(\tau) + K^{(2e)}_{\text{off}}(\tau) \notag \\ &
= \frac{\tau^3}{\beta^2} \left( \frac{1}{8} |{\cal D}|^4 
+ |{\cal D}|^2 \Im {\cal D} - \Re {\cal D}^2 + \frac{\beta}{2} 
|{\cal D}|^2 + (2 - \beta) |{\cal D}|^2 + 2 (2 - \beta)
\Im {\cal D} \right) \notag \\ & 
= \frac{\tau^3}{\beta^2} \left( \frac{1}{8} |{\cal D}|^4 
+ |{\cal D}|^2 \Im {\cal D} - \Re {\cal D}^2 +
|{\cal D}|^2 \right) \notag \\ & = 0 \; ,
\end{align}
which can be seen by using (\ref{did}). This shows that
off-diagonal terms cancel the diagonal term also in this order.
It implies that the form factor is determined by periodic orbits
alone, because the different terms which involve diffractive
orbits cancel each other. This might be true also for other
point-like sources of diffraction like e.\,g.\ Aharonov-Bohm flux
lines (see \cite{FK98}), although a quantitative analysis would
require here the use of uniform approximations.

\section{Universality in parametric correlations}

The cancellation of off-diagonal and diagonal terms
is conform with the expected universality of spectral
statistics in chaotic systems. Universality is, however,
not only expected in the properties of single systems,
but also in the way in which system properties vary when a
parameter of the system is changed. For example, random
matrix theory makes predictions about correlations between
densities of states for different parameter values.
The semiclassical calculation of diagonal and off-diagonal 
terms allows to test this prediction for systems with
a point-like scatterer where the parameter is the strength
of the scatterer.

In analogy to the spectral form factor, a parametric form
factor can be defined as Fourier transform of the parametric
two-point correlation function
\begin{equation} \label{koftaup}
K(\tau,x) = \int_{-\infty}^{\infty} \! \frac{\dd \eta}{\bar{d}(E)} \;
\left\langle d_{\text{osc}}\left( E + \frac{\eta}{2},
             X + \frac{x}{2} \right)
             d_{\text{osc}}\left( E - \frac{\eta}{2},
             X - \frac{x}{2} \right)
\right\rangle_E
\; \exp\left( 2 \pi i \eta \tau \bar{d}(E) \right) \; .
\end{equation}
Here $x$ is the parameter difference between two systems.
In order for this statistics to be universal the parameter
has to be chosen in a particular way. The requirement is
that the variance of the velocities, the derivatives of the
unfolded energies with respect to the parameter
($\partial \varepsilon_n/ \partial x$), is equal to unity.
The unfolded energies are obtained from the quantum energies
of the system by a scaling that leads to mean level
distance of one.

For random matrix ensembles, the parametric two-point
correlation function was derived in \cite{SA93b}
in the context of disordered metallic systems. 
For the GUE-result, which we discuss first, the
Fourier transform in (\ref{koftaup}) can be evaluated
in a closed form. It results in
\begin{equation} \label{kgue}
K^{\text{GUE}}(\tau,x) = \begin{cases}
\frac{\sinh(2 \pi^2 x^2 \tau^2)}{2 \pi^2 x^2 \tau} \,
\exp(-2 \pi^2 x^2 \tau) & \text{if $\tau < 1$,}\\
\frac{\sinh(2 \pi^2 x^2 \tau)}{2 \pi^2 x^2 \tau} \,
\exp(-2 \pi^2 x^2 \tau^2) & \text{if $\tau > 1$,}
\end{cases}
\end{equation}
and has for small values of $\tau$ the expansion
\begin{equation} \label{kgue2}
K^{\text{GUE}}(\tau,x) = 
\tau - 2 \pi^2 x^2 \tau^2 + 2 \pi^4 x^4 \tau^3 + \dots \; .
\end{equation}

We examine in the following whether the perturbation
by a point-like scatterer leads to universal correlations.
For large parameter differences $x$ the parametric
correlations for these systems cannot be expected
to be universal. The treatment of a delta-scatterer
by the method of self-adjoint extensions leads to a
quantisation condition with the property that there
is exactly one eigenvalue of the perturbed system
within each pair of neighbouring eigenvalues of the
unperturbed system. This puts a restriction
to the movement of eigenvalues when the parameter is
changed. For this reason, one can expect universal
properties only for small parameter differences.

We choose first $a$ in (\ref{d}) as parameter
of the system. The two densities in (\ref{koftaup})
differ then only in the diffraction coefficient
${\cal D}$. As a consequence, the results for
the parametric form factor can be obtained
directly from the spectral form factor without
further calculations. One has to express the
contributions to the form factor, (\ref{kd1})
and (\ref{kd2}), in terms of ${\cal D}_1$ and
${\cal D}_1^*$ and replace ${\cal D}_1^*$ by ${\cal D}_2^*$.
For $\beta = 2$ one obtains in this way
\begin{align} \label{kparasc}
\tilde{K}_{\text{sc}}(\tau,x) - \tilde{K}_{\text{sc}}(\tau,0) 
& \approx \frac{\tau^2}{4} \left[
{\cal D}_1 {\cal D}_2^* - 2 i \, {\cal D}_1 + 2 i \, {\cal D}_2^* \right]
+ \frac{\tau^3}{32} \left[ ({\cal D}_1 {\cal D}_2^*)^2
\right. \notag \\ & \qquad \qquad \left. 
- 4 i \, {\cal D}_1 {\cal D}_2^* ({\cal D}_1 
- {\cal D}_2^*) - 4
{\cal D}_1 {\cal D}_1 - 4 {\cal D}_2^* {\cal D}_2^*
+ 8 {\cal D}_1 {\cal D}_2^* \right] \notag \\ &
= \frac{\tau^2}{4} \left[ {\cal D}_1 {\cal D}_2^* - 
2 i \, {\cal D}_1 + 2 i \, {\cal D}_2^* \right]
+ \frac{\tau^3}{32} \left[ {\cal D}_1 {\cal D}_2^* - 
2 i \, {\cal D}_1 + 2 i \, {\cal D}_2^* \right]^2 \; .
\end{align}
A first point to notice is that the parametric form factor
in (\ref{kparasc}) is, in general, not real. This seems to be
in contrast to the random matrix result (\ref{kgue}) which 
is real. The reason for this lies, however, in the correct
choice of the unfolding procedure. The definition (\ref{koftaup})
yields only the universal form factor in case
that the mean density of states $\bar{d}(E)$
does not depend on the parameter of the system. However,
in the present case the mean density changes slightly
with the parameter $x$ of the system, and this leads
to a slight shift of the spectrum with $x$ \cite{Sie00}.
As a consequence, the argument $\eta$ of the two-level
correlation function is shifted, and its Fourier transform,
the form factor, is multiplied by a term of the form
$e^{i c \tau}$, where $c$ is determined by the shift
of the levels. By rewriting (\ref{kparasc})
up to the considered order in $\tau$ in the form
\begin{align} \label{kparasc2}
\tilde{K}_{\text{sc}}(\tau,x) - \tilde{K}_{\text{sc}}(\tau,0) 
& \approx \left( \tau + \frac{\tau^2}{4} \Re \left[ {\cal D}_1
{\cal D}_2^* - 2 i \, {\cal D}_1 + 2 i \, {\cal D}_2^* \right]
+ \frac{\tau^3}{32} \left( \Re \left[ {\cal D}_1 {\cal D}_2^* - 
2 i \, {\cal D}_1 + 2 i \, {\cal D}_2^* \right] \right)^2
\right) \notag \\ & \qquad \qquad
\times \exp \left( \frac{i \tau}{4} \Im \left[ {\cal D}_1
{\cal D}_2^* - 2 i \, {\cal D}_1 + 2 i \, {\cal D}_2^* \right]
\right)  - \tau \; ,
\end{align}
one can extract the result that corresponds to
a proper unfolding by dropping the exponential,
and one obtains
\begin{align} \label{kparasc3}
K_{\text{sc}}(\tau,x) - K_{\text{sc}}(\tau,0) 
& \approx \frac{\tau^2}{4} \Re \left[ {\cal D}_1
{\cal D}_2^* - 2 i \, {\cal D}_1 + 2 i \, {\cal D}_2^* \right]
+ \frac{\tau^3}{32} \left( \Re \left[ {\cal D}_1 {\cal D}_2^* - 
2 i \, {\cal D}_1 + 2 i \, {\cal D}_2^* \right] \right)^2
\notag \\ & 
= - \frac{\tau^2}{8} \, |{\cal D}_1 - {\cal D}_2|^2
+ \frac{\tau^3}{128} \, |{\cal D}_1 - {\cal D}_2|^4 \; ,
\end{align}
where (\ref{did}) has been used.

A comparison with (\ref{kgue2}) shows that it has now
the same form as the random matrix result. The remaining
step is to connect the
parameter $a$ with the universal parameter $x$. Since
we consider small parameter differences, $x$ is
given by $x = \Delta a \, \sigma_v$, where $\sigma_v$
is the square root of the variance of the velocities
with respect to the parameter $a$. In order to evaluate
$\sigma_v$ we employ a semiclassical method
\cite{EFKAMM95,CKR98,LS99} which expresses it in the form
\begin{equation} \label{sigv}
\sigma_v^2 = \left\langle \left( \frac{\partial \varepsilon_n}{
\partial a} \right)^2 \right\rangle_E
= \left\langle \lim_{\eta \rightarrow 0} 2 \pi \eta
\left[ d_v^\eta (\varepsilon) \right]^2 \right\rangle_E \; ,
\end{equation}
where $d_v^\eta$ is a Lorentzian smoothed density of 
states, that is weighted by the velocities and expressed
semiclassically by
\begin{equation} \label{dv}
d_v^\eta (\varepsilon) = \sum_n \frac{\partial \varepsilon_n}{
\partial a} \, 
\frac{1}{\pi} \, \frac{\eta}{(\varepsilon - \varepsilon_n)^2 
+ \eta^2}
\approx \Im \frac{\partial}{\partial a}
\sum_\gamma \frac{A_\gamma}{\pi T_\gamma}
\exp\left\{ \frac{i}{\hbar} S_\gamma - 
\frac{\eta T_\gamma}{\hbar \bar{d}(E)} \right\} \; .
\end{equation}
Here $\eta$ is the width of the Lorentzian and all
other quantities are defined as before.

For small parameter differences we can express the derivative
of the diffraction coefficient with respect to $a$ by
$({\cal D}_2 - {\cal D}_1)/\Delta a$. After inserting
(\ref{dv}) into (\ref{sigv}) the leading order contribution
comes from single-diffractive orbits, and we evaluate
the double sum in the diagonal approximation with the
sum rule (\ref{sum1}) and amplitudes (\ref{a1do})
\begin{align}
\sigma_v^2 
& = \lim_{\eta \rightarrow 0} \int_0^\infty \! \dd T \;
\sum_\gamma \frac{\eta \hbar \delta(T - T_\gamma)}{
4 \pi^2 p^2_v \beta |M_{12}|} \, \frac{|{\cal D}_2 - {\cal D}_1|^2}{
(\Delta a)^2} \exp\left\{ - 
\frac{2 \eta T_\gamma}{\hbar \bar{d}(E)} \right\}
\notag \\ & =  \lim_{\eta \rightarrow 0} \int_0^\infty \! \dd T \;
\frac{\eta \hbar}{\beta \Sigma(E)} \, \frac{|{\cal D}_2 - {\cal D}_1|^2}{
(\Delta a)^2} \exp\left\{ - 
\frac{2 \eta T}{\hbar \bar{d}(E)} \right\}
\notag \\ & = \frac{2}{\beta} \, \frac{|{\cal D}_2 - {\cal D}_1|^2}{
(4 \pi \Delta a)^2} \; ,
\end{align}
For $\beta=2$, the universal parameter is given by
$x = \Delta a \, \sigma_v = |{\cal D}_2 - {\cal D}_1|/(4 \pi)$,
and after insertion into (\ref{kparasc3}) we reproduce the
random matrix result (\ref{kgue2})
\begin{equation} \label{kparasc4}
K_{sc}(\tau,x) - K_{sc}(\tau,0) \approx 
- 2 \pi^2 x^2 \tau^2 + 2 \pi^4 x^4 \tau^3 \; .
\end{equation}

For the GOE ensemble, the parametric density correlation
function is given by a triple-integral which cannot be
expressed in closed form \cite{SA93b}. We consider here
only the first-order correction for which the GOE-result
can be obtained from the asymptotic form of the correlation
function for long-range correlations in \cite{BK96,OLM98}:
$K^{\text{GOE}}(\tau,x) - K^{\text{GOE}}(\tau,0) \approx
- 2 \pi^2 x^2 \tau^2$. 
Again we find an agreement
with the semiclassical result 
$K_{sc}(\tau,x) - K_{sc}(\tau,0)
\approx - \tau^2 \, |{\cal D}_1 - {\cal D}_2|^2/4
= - 2 \pi^2 x^2 \tau^2$.

\section{Discussion}

We have investigated in this article the influence of a
point-like scatterer on the spectral statistics of quantum
systems with chaotic classical limit. It has been shown
that the modification of the form factor $K(\tau)$ 
due to the scatterer can be
evaluated systematically in a semiclassical approximation.
The expansion of the form factor in powers of $\tau$
corresponds on the semiclassical side to an expansion
in the number of loops of the diffractive orbits. 
We have calculated off-diagonal contributions to the
$\tau^2$- and $\tau^3$-term, but the method can be
extended to higher order terms.

The results lead to the conclusion that the
delta-perturbation does not modify the form factor.
Off-diagonal terms from pairs of different
diffractive orbits and from pairs of diffractive
and periodic orbits cancel exactly the diagonal
terms from diffractive orbits. This requires the
existence of correlations between different orbits.
These correlations arise from pairs of orbits
which are very close in coordinate space. 

The results provide a support for the
random-matrix conjecture. They imply, up
to the considered order, that the statistics of
chaotic systems are invariant under the perturbation
by a point-like scatterer as is expected from
universality. They show also that
correlations between two energy spectra for 
different parameter values are universal,
provided that the parameter difference is small.
Furthermore, they indicate indirectly that the
spectral statistics of the unperturbed system
(and thus also of the perturbed system) are identical
with those of random matrix theory. The reason for
this is that independent results on the invariance
of spectral statistics under a delta-perturbation
are based on the assumption that the
unperturbed energy levels and wave functions
have random matrix distributions \cite{SS98b,BLS00}.
Since the semiclassical results show this
invariance for chaotic systems, the combination
of both results provides a theoretical
indication that chaotic systems follow the
random matrix hypothesis.

Finally, the results are a support for the
semiclassical method. They show that semiclassical
approximations are capable to go beyond the
leading term in $\tau$ and are an appropriate
tool for investigating spectral statistics
in the semiclassical limit.

\bigskip \bigskip

\noindent
I would like to thank K.\ Richter and P.\ \v{S}eba
for helpful discussions. After completion of this
article I learned about work by E.\ Bogomolny, P.\ Leboeuf
and C.\ Schmit \cite{BLS00} with related semiclassical
results for the first-order correction.

\appendix

\section{Stability matrix for the motion in a magnetic field}

The stability matrix $\tilde{M}$ of a trajectory
determines infinitesimal orthogonal deviations from the
final point of the trajectory in terms of the deviations
from the initial point of the trajectory
\begin{equation}
\begin{pmatrix} \dd r_f \\ \dd p_f \end{pmatrix} = \tilde{M}
\begin{pmatrix} \dd r_i \\ \dd p_i \end{pmatrix} \; .
\end{equation}
In systems with a magnetic field the momentum has the
form $\boldsymbol{p} = m \boldsymbol{v} + \frac{q}{c}
\boldsymbol{A}(\boldsymbol{r})$,
where $q$ is the charge of the particle and $\boldsymbol{A}$ is
the vector potential. In this case it is often more convenient
to consider a matrix $M$ that describes deviations
of the velocities instead of those of the momenta.
The relation between both matrices is given by
\begin{equation}
\begin{pmatrix} \dd r_f \\ m \dd v_f \end{pmatrix} = M
\begin{pmatrix} \dd r_i \\ m \dd v_i \end{pmatrix}
\; , \qquad M = A_f^{-1} \tilde{M} A_i \; , \qquad
A_{i,f} = \begin{pmatrix} 1 & 0 \\ a_{i,f} & 1 \end{pmatrix} \; ,
\end{equation}
where $a_{i,f} = \frac{q}{c} (\hat{n}_{i,f} \cdot \nabla) (
\hat{n}_{i,f} \cdot \boldsymbol{A}(\boldsymbol{r}_{i,f}))$,
and $\hat{n}_{i,f}$ is the direction orthogonal
to the trajectory at the initial and final point of the
trajectory, respectively. The matrix $M$ has unit determinant
and satisfies $M_{12} = \tilde{M}_{12}$
and $\Tr M = \Tr \tilde{M} + \tilde{M}_{12} (a_i - a_f)$.
In cases where the initial and final points are identical
and the initial and final velocities differ by a small angle
$\varepsilon$, the traces are identical in leading order
of $\varepsilon$. Since the semiclassical approximations
in this article involve only $\tilde{M}_{12}$ and
$\Tr \tilde{M} - 2$, we express all quantities 
in terms of $M$ instead of $\tilde{M}$, and we use
also the term stability matrix for it.

\end{document}